\documentclass[twocolumn,showpacs,preprintnumbers,amsmath,amssymb,superscriptaddress]{revtex4}

\usepackage{graphicx,bm}

\newcommand{\Sec}[1]{Sec.\,\ref{#1}}

\newcommand{\be}{\begin{equation}}
\newcommand{\ee}{\end{equation}}
\newcommand{\bea}{\begin{eqnarray}}
\newcommand{\eea}{\end{eqnarray}}
\newcommand{\bsube}{\begin{subequations}}
\newcommand{\esube}{\end{subequations}}
\newcommand{\Fig}[1]{Fig.\,\ref{#1}}
\newcommand{\Eq}[1]{Eq.\,(\ref{#1})}

\newcommand{\alf}{\alpha}
\newcommand{\sgm}{\sigma}
\newcommand{\omg}{\omega}
\newcommand{\Omg}{\Omega}
\newcommand{\Gam}{\Gamma}

\newcommand{\vpl}{\varepsilon}
\newcommand{\epl}{\epsilon}
\newcommand{\Dlt}{\Delta}
\newcommand{\dlt}{\delta}

\newcommand{\la}{\langle}
\newcommand{\ra}{\rangle}
\newcommand{\ti}{\tilde}
\newcommand{\rmd}{{\rm d}}

\newcommand{\rmw}{{\rm w}}

\newcommand{\rmi}{{\rm i}}
\newcommand{\tr}{{\rm tr}}

\begin{document}

\title{Non-Markovian dynamics and noise characteristics in continuous
measurement of a solid--state charge qubit}

 \author{JunYan Luo}
 \email{jyluo@zust.edu.cn}
 \affiliation{School of Science, Zhejiang University of Science
  and Technology, Hangzhou 310023, China}

 \author{HuJun Jiao}
 \affiliation{Department of Physics, Shanxi University, Taiyuan,
 Shanxi 030006, China}

 \author{Xiao-Li Lang}
 \affiliation{School of Science, Zhejiang University of Science
  and Technology, Hangzhou 310023, China}

 \author{BiTao Xiong}
 \affiliation{School of Science, Zhejiang University of Science
  and Technology, Hangzhou 310023, China}

 \author{Xiao-Ling He}
 \affiliation{School of Science, Zhejiang University of Science
  and Technology, Hangzhou 310023, China}

\begin{abstract}
 We investigate the non-Markovian characteristics in continuous
 measurement of a charge qubit by a quantum point contact.
 The backflow of information from the
 reservoir to the system in the non-Markovian domain gives
 rise to strikingly different qubit relaxation and
 dephasing in comparison with the Markovian case.
 The intriguing non-Markovian dynamics is found to have a direct
 impact on the output noise feature of the detector.
 Unambiguously, we observe that the non-Markovian
 memory effect results in an enhancement of the signal-to-noise
 ratio, which can even exceed the upper limit of ``4'', leading
 thus to the violation of the Korotkov-Averin bound in quantum
 measurement.
 Our study thus may open new possibilities to improve detector's
 measurement efficiency in a direct and transparent way.
 \end{abstract}



\maketitle

\section{\label{thsec1}Introduction}

 Quantum coherent oscillations in a quantum two-level system
 (qubit) stand for the most basic dynamic manifestation of
 quantum coherence between the qubit states.
 Motivated by potential applications to quantum
 computation \cite{Bra92,All11072138}, as well as
 general interest in mesoscopic quantum phenomena,
 intensive experimental and theoretical effort has been
 devoted to the attempts to study these oscillations
 and measurement possibilities of individual qubits.
 One of the especially interesting methods of detecting
 coherent oscillations is to monitor them continuously with a
 mesoscopic electrometer, such as single electron transistor
 (SET) \cite{Shn9815400,Dev001039,Mak01357,Cle02176804,%
 Jia07155333,Gil06116806,Gur05073303,Oxt06045328} or
 quantum point contact (QPC)
 \cite{Buk98871,Sch981238,Nak99786,Spr005820,Aas013376},
 whose conductance depends on the charge state of a nearby
 qubit.
 From the readout of the detector, one is capable of gaining
 essential insight into the nontrivial correlation
 characteristics between the detector and the measured
 system.

 In contrast to the projective measurement which takes
 place instantaneously, the continuous detection extracts
 information of the measured system continually.
 However, the detection inevitably acts back on the
 system, leading thus to the dephasing of the qubit.
 This trade-off between acquisition of quantum state information
 and backaction dephasing of the measured system plays the central
 role in the process of quantum measurement.
 Recently, it was demonstrated that the measurement properties
 are intimately associated with the full counting statistics
 of the detector \cite{Naz03,Ave05126803}.
 Evaluation of the shot noise and higher order cumulants
 of quantum measurement have been worked out under Markovian
 approximation and in the wide-band limit
 (WBL) \cite{Rod05251,Moz04018303,Cle04121303,Wab05165347,%
 Goa01125326,Kor01115403,Sta04136802,Fli04205334,Kie06033312,Wan07125416}.
 The Markovian approximation assumes that the correlation time in the reservoirs
 is much shorter than the typical response time in the reduced
 system, while the WBL neglects the energy-dependent densities of
 states in the electrodes.
 Yet, these approximations may not be always true in realistic
 devices.
 Hence, a recent development in noise characteristics has
 been devoted to the investigations of the non-Markovian
 effects with energy-dependent spectral density in the
 environment \cite{Fen08075302,Zhe09124508,Zed09045309}.
 Flindt et al \cite{Fli08150601} and Aguado et al \cite{Agu04206601}
 studied the noise properties of a charge qubit in a transport
 configuration, where the non-Markovian effect of the phonon bath
 was effectively accounted for.
 The non--Markovian correlations of electrodes were investigated
 in the context of transport through quantum dot (QD) systems
 \cite{Jin11053704,Bra06026805}, and measurement of a nanomechanical
 resonator by QPC \cite{Che11012393}, where radical difference
 in dynamics between non-Markovian and Markovian cases was
 identified.

 The purpose of this paper is to study the non--Markovian
 characteristics in continuous measurement of a qubit by a QPC.
 Our analysis is based on a generalized time-nonlocal
 quantum master equation approach \cite{Yan982721},
 which is capable of treating properly the energy exchange
 between the qubit and detector.
 In comparison with the Markovian case, we find considerable
 differences in qubit relaxation and dephasing in the
 non-Markovian domain where the information may flow
 from the environment back to the reduced system.
 Furthermore, the unique non-Markovian dynamics
 is reflected in the output noise feature of the detector.
 We observe that the non-Markovian memory
 effect results in a strong enhancement of the
 signal-to-noise ratio (SNR).
 It is demonstrated unambiguously that under appropriate
 conditions the SNR can even exceed the upper limit of
 ``4'', leading thus to the violation of the Korotkov-Averin
 (K-A) bound.
 Our study thus may open new possibilities to improve detector's
 measurement efficiency in a direct and transparent way.

 This paper is organized as follows. We begin in \Sec{thsec2} with the
 model set-up of a charge qubit under the continuous monitoring by a
 QPC. In \Sec{thsec3}, we first analyze the reservoir correlation time
 under various parameters of the QPC detector, and then study the unique
 qubit relaxation and dephasing arising from the non-Markovian processes.
 \Sec{thsec4} is devoted to the calculation of noise characteristics
 based on the ``$N$''-resolved time-nonlocal quantum master equation
 approach.
 Numerical results, particularly, the discussions of SNR under various
 parameters are presented.
 Finally, we summarize the main results and implications of this work
 in \Sec{thsec5}.

 \begin{figure}
 \begin{center}
 \includegraphics*[scale=0.8]{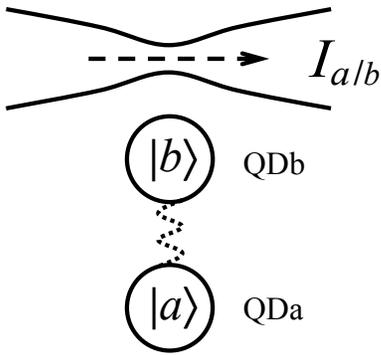}
 \caption{\label{fig1}
 Schematic setup of a solid--state charge qubit
 measured continuously by a quantum point contact.
 The qubit is represented by an extra electron
 tunneling between the coupled quantum dots.
 The tunneling amplitude of the QPC is susceptible to
 changes in the surrounding electrostatic environment,
 and can therefore be used to sense the position
 of the extra electron.}
 \end{center}
 \end{figure}

 \section{\label{thsec2}Model description}

 The system under investigation is schematically shown
 in \Fig{fig1}. The qubit is represented by an extra
 electron tunneling between two coupled quantum dots (QDa and QDb).
 When the electron occupies the QDa (QDb), the qubit
 is said to be in the localized state $|a\ra$ ($|b\ra$).
 A nearby QPC serves as the charge detector to continuously
 monitor the position of the electron.
 Occupation of the electron in different dots
 leads to distinct influence on the transport current
 through the QPC. It is right this mechanism that makes
 it possible to read out the qubit-state information.
 The entire system is described by the Hamiltonian
 $H_{\rm T} =H_{\rm qu}+H_{\rm D}+H'$, where
 \bsube\label{sys-Ham}
 \begin{gather}
 H_{\rm qu} =\frac{1}{2}\epl\sgm_z+\Omg\sgm_x,\label{Hqubit}
 \\
 H_{\rm D} =\sum_{k\in {\rm L}} \vpl_k
 \hat c_k^\dag \hat c_k
 +\sum_{q\in {\rm R}} \vpl_q
 \hat c_q^\dag \hat c_q \, ,
 \\
 H' =\sum_{s=a,b}\sum_{k,q}t^{s}_{kq}
 \hat c_k^\dag \hat c_q
 \cdot |s\ra\la s|+{\rm h.c.}.\label{Hprim}
 \end{gather}
 \esube
 Here, $H_{\rm qu}$ denotes the qubit Hamiltonian, where
 the pseudospin operators are defined as
 $\sgm_z\equiv|a\ra\la a|-|b\ra\la b|$ and
 $\sgm_x\equiv|a\ra\la b|+|b\ra\la a|$, respectively.
 Each dot has only one bound state, i.e., the logic
 states $|a\ra$ and $|b\ra$, with level detuning $\epl$
 and interdot coupling $\Omg$.

 The second component $H_{\rm D}$ depicts the left and right
 QPC reservoirs, where $\hat{c}_k$ ($\hat{c}_q$) denotes
 the annihilation operator for an electron in the left (right)
 QPC reservoir.
 The electron reservoirs are characterized by the Fermi
 distributions $f_\alf{(\omg)}=\{1+e^{\beta(\omg-\mu_\alf)}\}^{-1}$,
 where $\mu_\alf$ is the Fermi energy of the left ($\alf$=L)
 or right ($\alf$=R) reservoir, and $\beta=(k_{\rm B}T)^{-1}$ is the
 inverse temperature.
 Hereafter, the Planck's constant $\hbar$ and the electron charge $e$
 are set to unity, i.e. $\hbar=e=1$, unless stated otherwise.
 Throughout this work, we define
 $\mu^{\rm eq}_{\rm L}=\mu^{\rm eq}_{\rm R}=0$ for the
 equilibrium chemical potentials  (or Fermi energies)
 of the QPC reservoirs.
 An applied measurement voltage thus is modeled by the difference
 in chemical potentials of the left and right electrodes:
 $V=\mu_{\rm L}-\mu_{\rm R}$.

 The tunneling Hamiltonian for the QPC detector is
 represented by the last component $H'$.
 The amplitude $t^s_{kq}$ of electron tunneling
 through two reservoirs of the QPC depends explicitly
 on the qubit state $|s\ra$ ($s=a$ or $b$).
 Thus the quantum operator to be measured is $\sgm_z$.
 By denoting $Q_s\equiv |s\ra\la s|$, the qubit--QPC detector
 coupling in the $H_{\rm D}$--interaction picture can be rewritten
 as $H'(t)=\sum_s[\hat f_s(t)+\hat f^{\dag}_s(t)]\cdot Q_s$,
 with
 $\hat f_s(t)\equiv e^{\rmi H_{\rm D}t}\big(\sum_{kq}t_{kq}^s
 \hat c_k^\dag \hat c_q\big)e^{-\rmi H_{\rm D}t}$.
 The effects of the stochastic QPC reservoirs on
 measurement are characterized by the reservoir correlation
 functions
 $C_{ss'}^{(+)}(t-\tau)\equiv\la \hat f_{s}^\dag(t)
 \hat f_{s'}(\tau)\ra$ and
 $C_{ss'}^{(-)}(t-\tau)\equiv\la \hat f_{s}(t)
 \hat f_{s'}^\dag(\tau)\ra$.
 By introducing the reservoir spectral density function
 \be\label{Jww}
 J_{ss'}(\omg,\omg')=\sum_{k,q} t_{kq}^st_{kq}^{s'}
 \delta(\omg-\vpl_k)\delta(\omg'-\vpl_q),
 \ee
 these QPC coupling correlation functions can be recast as
 \be\label{CCF}
 C_{ss'}^{(\pm)}(t) = \! \int\!\!\!\int\!\! \rmd\omg \rmd\omg'
 J_{ss'}(\omg,\omg')f_{\rm L}^{(\pm)}(\omg)
 f_{\rm R}^{(\mp)}(\omg') e^{\pm \rmi(\omg-\omg')t},
 \ee
 where $f_\alf^{(+)}(\omg)$ is the usual Fermi function, and
 $f_\alf^{(-)}(\omg)\equiv1-f_\alf^{(+)}(\omg)$.

 In order to characterize finite cutoff energy of the QPC reservoirs,
 we introduce a single Lorentzian to model the band structure.
 For the sake of constructing analytical results, we assume
 a simple Lorentzian function of cutoff ``w'' centered at
 the Fermi energy for the spectral density \Eq{Jww}.
 Moreover, the bias voltage is conventionally described by
 a relative shift of the entire energy-bands, thus the
 centers of the Lorentzian functions would fix at the
 Fermi levels $\mu_{\rm L}$ and $\mu_{\rm R}$. Without loss of
 generality, we set \cite{Luo09385801}
 \be\label{Jw}
 J_{ss'}(\omg,\omg')=\chi_s\chi_{s'}
 \frac{\Gam_{\rm L}^0\rmw^2}{(\omg-\mu_{\rm L})^2+\rmw^2}\cdot
 \frac{\Gam_{\rm R}^0\rmw^2}{(\omg'-\mu_{\rm R})^2+\rmw^2},
 \ee
 where $\Gam_{\rm L}^0$ $(\Gam_{\rm R}^0)$ is the maximum
 of the spectral in the left (right) electrode;
 $\chi_s$ and $\chi_{s'}$ are qubit-QPC coupling
 parameters, which are of $\chi_a>\chi_b$, as can be inferred
 from \Fig{fig1}.
 In the limit of w$\rightarrow\infty$,  the QPC spectral density
 \Eq{Jw} becomes energy-independent and reduces to the constant
 WBL spectral density used in the literature.

 \section{\label{thsec3}Non-Markovian Dynamics}

 The non--Markovian dynamics of the reduced system is described
 by a generalized time-nonlocal quantum master equation.
 An equation of this type can be obtained from the partitioning
 scheme devised by Nakajima and Zwanzig \cite{Zwa01,Bre02},
 or in the real-time diagrammatic technique for the dynamics
 of the reduced density matrix on the Keldysh contour \cite{Mak01357},
 \be\label{QME}
 \dot{\rho}(t)=-\rmi {\cal L}\rho(t)-\!\int_{-\infty}^t\! \rmd \tau
 \Pi(t-\tau)\rho(\tau),
 \ee
 where the first term ${\cal L}(\cdots)\equiv[H_{\rm qu},(\cdots)]$ is the
 qubit Liouvillian.
 The influence of the QPC detector on the dynamics of the qubit
 is described by the memory kernel (the second term), which is
 given by \cite{Yan982721},
 \begin{align}
 \!\!\Pi(t-\tau)\rho(\tau)=
 \sum_{ss'}[&Q_s,C_{ss'}(t-\tau)G(t-\tau)Q_{s'}\rho(\tau)
 \nonumber \\
 &\;\;-C^\ast_{ss'}(t-\tau)G(t-\tau)\rho(\tau)Q_{s'}],
 \end{align}
 where $C_{ss'}(t-\tau)=C^{(+)}_{ss'}(t-\tau)+C^{(-)}_{ss'}(t-\tau)$,
 and $G(t-\tau)\equiv e^{-{\rmi\cal L}(t-\tau)}$ is the
 free propagator associated with the qubit Hamiltonian alone.
 In deriving \Eq{QME}, the only approximation made is the
 second-order perturbation in the system-reservoir coupling.
 This equation thus is valid for arbitrary reservoir temperatures,
 cutoff energies, and measurement voltages, as long as the
 second-order perturbation holds.

 \begin{figure}
 \begin{center}
 \includegraphics*[scale=0.72]{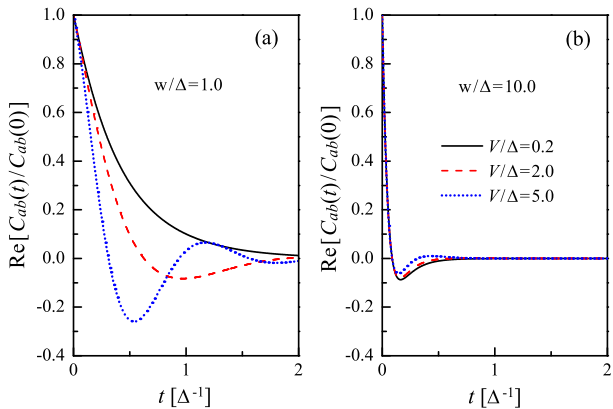}
 \caption{\label{fig2}
 Real part of reservoir correlation function $C_{ab}(t)$ at
 (a) a small cutoff energy w$/\Dlt=1.0$ and (b) a large
 cutoff energy w$/\Dlt=10.0$ for different bias voltages:
 $V/\Dlt=0.2$ (solid lines), $V/\Dlt=2.0$ (dashed lines),
 and $V/\Dlt=5.0$ (dotted lines).
 The temperature is $\beta\Dlt=1.0$.}
 \end{center}
 \end{figure}

 The Markovian approximation is valid only when the correlation
 time of the reservoir is much shorter than the characteristic
 time of the reduced quantum system.
 The former one is defined as the time scale at which the
 profiles of the QPC reservoir two-time correlation function
 decays.
 It is closely associated with the following time scales,
 the time scale of the QPC spectral density
 ($\sim$w$^{-1}$), the time scale of the applied bias ($\sim V^{-1})$,
 and the time scale of the QPC reservoir temperature ($\sim\beta$).
 In what follows, we first study numerically how the reservoir
 correlation time is varied as a function of different parameters of the
 QPC reservoirs.

 \Fig{fig2} shows the real parts of the QPC reservoir correlation
 function $C_{ab}(t)$ versus
 time for various values of cutoff energy and voltage at a given
 temperature $\beta\Dlt=1.0$.
 The correlation time decreases as the voltage increases, which is
 particularly prominent for a narrow cutoff w$/\Dlt=1.0$
 as displayed in \Fig{fig2}(a).
 By comparing \Fig{fig2} (a) and (b), it is revealed that
 the dependence of reservoir correlation time on the bias
 voltage is much weaker than that on the cutoff.
 In the limit of w$\rightarrow0$, one finds
 $J_{ss'}(\omg,\omg')\propto\dlt(\omg-\mu_{\rm L})\dlt(\omg'-\mu_{\rm R})$, which
 leads to QPC correlation
 functions proportional to $e^{\pm\rmi Vt}$, i.e. completely
 non-local in time.
 The opposite limit of WBL (w$\rightarrow\infty$) corresponds to a channel-mixture
 regime, where a great number of possible transitions of electron
 tunneling between the two reservoirs of the QPC
 take place.
 It was shown in this regime the QPC reservoir correlation time
 and memory effect are remarkably reduced \cite{Lee08224106}.
 Thus, the larger the cutoff is, the shorter the
 correlation time is. The reservoir correlation time is
 mainly restricted by the cutoff.

 \begin{figure}
 \begin{center}
 \includegraphics*[scale=0.72]{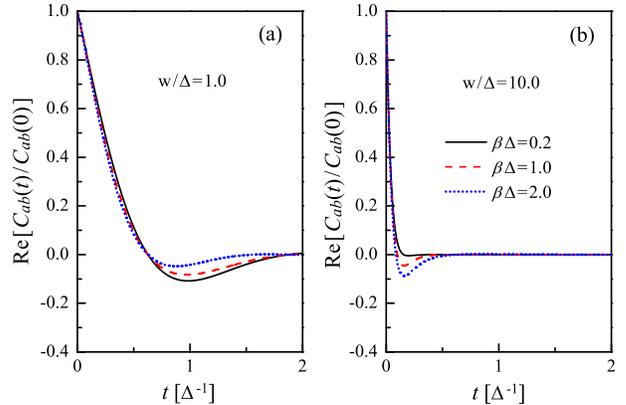}
 \caption{\label{fig3}
 Real part of reservoir correlation function $C_{ab}(t)$ vs time
 ``$t$'' for (a) w$/\Dlt=10.0$ and (b) w$/\Dlt=1.0$
 at different temperatures: $\beta\Dlt=0.2$ (solid lines), $\beta\Dlt=1.0$ (dashed
 lines), and $\beta\Dlt=2.0$ (dotted lines).
 The measurement voltage is $V/\Dlt=2.0$.}
 \end{center}
 \end{figure}

 To explore the influence of the temperature, we plot in
 \Fig{fig3} the reservoir correlation function for various
 temperatures.
 Analogous to that on the measurement voltage, the correlation
 time deceases as the reservoir temperature rises.
 Nevertheless, the effect of the temperature on the reservoir
 correlation time is less sensitive than that on the voltage,
 as can be seen by comparing \Fig{fig2} and \Fig{fig3}.
 Therefore, the cutoff energy has the dominant role to play in 
 determining the QPC reservoir correlation time.

 \begin{figure}
 \begin{center}
 \includegraphics*[scale=0.8]{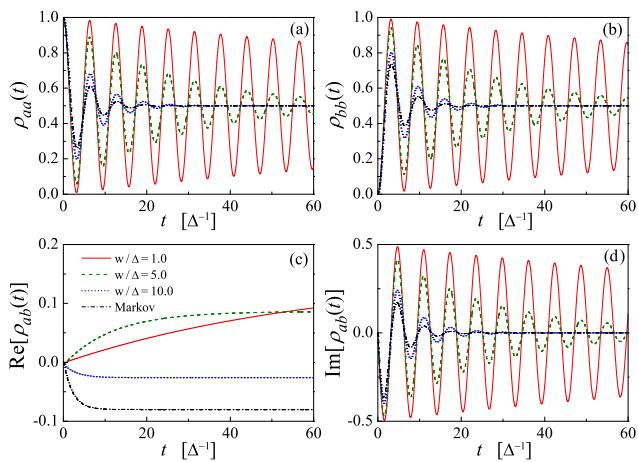}
 \caption{\label{fig4}
 Non-Markovian dynamics of the qubit for different values of
 cutoff energy: w$/\Dlt=1.0$ (solid lines), w$/\Dlt=5.0$ (dashed
 lines), and w$/\Dlt=10.0$ (dotted lines).
 (a) The probability of finding the electron in the localized state $|a\ra$:
 $\rho_{aa}(t)\equiv\la a|\rho(t)|a\ra$,
 (b) in the  localized state $|b\ra$: $\rho_{bb}(t)\equiv\la b|\rho(t)|b\ra$,
 (c) real part of the off-diagonal matrix element $\rho_{ab}(t)$, and
 (d) imaginary part of $\rho_{ab}(t)$.
 The Markovian WBL results are also plotted in dash-dotted lines
 for comparison.
 The qubit is assumed to be symmetric ($\epl=0$), and
 initially in the state $\rho_{\rm ini}=|a\ra\la a|$.
 Other parameters used are $V/\Dlt=5.0$, $\beta\Dlt=1.0$,
 $\eta\Dlt^2=2.0$, and the tunneling amplitudes $\chi_a/\Dlt=1.0$
 and $\chi_b/\Dlt=0.8$.}
 \end{center}
 \end{figure}

 With the knowledge of these time scales, we are now
 in a position to discuss the non-Markovian dynamics of the
 qubit under the continuous measurement of the QPC.
 The numerical propagation of the time-nonlocal QME
 is facilitated by employing the approach of auxiliary density
 operators \cite{Wel06044712,Jin08234703,Cro09073102,Cro10159904}.
 The calculation of the time evolution is then reduced
 to the propagation of coupled differential equations.
 The numerical results of the non-Markovian dynamics
 of the qubit are plotted in \Fig{fig4}
 for different values of the cutoff energy.
 For comparison, we have also plotted the Markovian
 result by the dash-dotted lines for the same parameters.

 The measurement backaction-induced dephasing leads to
 a coherent to incoherent transition of the qubit
 electron tunneling.
 In the coherent regime, the tunneling leads to
 the well-known Rabi oscillations with frequency
 given by $\Dlt$, as indicated in \Fig{fig4}(a)
 and (b).
 For a symmetric qubit ($\epl=0$), the occupation
 probability in each dot finally reaches 1/2 for
 both Markovian and non-Markovian cases.
 However, for a small cutoff, such as w$/\Dlt=1.0$ (solid lines),
 the non-Markovian relaxation behavior shows a considerable
 difference to the Markovian case (dash-dotted lines).
 As the cutoff energy increases, the qubit relaxation gets close
 to that of the Markovian result, due to reduced reservoir
 correlation time [see the dotted lines in
 \Fig{fig4}(a) and (b)].

 The backaction-induced dephasing behavior is described
 by the off-diagonal density-matrix element, as displayed
 in \Fig{fig4}(c) and (d).
 The real part of $\rho_{ab}$ approaches a nonzero
 constant at long times.
 The nonzero stationary result stems from the energy exchange
 between the qubit and QPC detector \cite{Li04085315,Li05066803,Luo09385801}.
 For both Markovian and non-Markovian cases, the imaginary
 part of $\rho_{ab}$ goes to zero in the stationary limit.
 However, for a small cutoff energy, the dephasing
 rate is much lower than that of the Markovian case, as displayed
 by the solid line in \Fig{fig4}(d).
 In Markovian processes, information flows continuously
 from the qubit to its environment.
 Yet, in the presence of non-Markovian behavior, a reversed
 flow of information from the environment
 back to the reduced system occurs, which
 leads to the reduction of the dephasing rate.

 To clearly demonstrate this unique feature, we
 employ the ``trace distance'' of two quantum states $\rho_1$ and
 $\rho_2$, which is defined as \cite{Nie00,Bre09210401,Lai10062115}
 \be
 D[\rho_1(t),\rho_2(t)]=\frac{1}{2}{\rm tr}|\rho_1(t)-\rho_2(t)|.
 \ee
 Here the norm is given by $|A|=\sqrt{A^\dag A}$, and
 $\rho_{1,2}(t)$ are the dynamical qubit states for a given
 pair of initial states $\rho_{1,2}(0)$.
 The trace distance describes the probability of distinguishing
 those states.
 In Markovian processes, the distinguishability between
 any two states are continuously reduced, and thus the trace
 distance $D(\rho_1,\rho_2)$ decreases monotonically.
 The essential property of non-Markovian behavior is
 the growth of this distinguishability.
 An increase of the trace distance during any time
 intervals implies the emergence of non-Markovianity
 (inverse flow of the information).
 One is therefore inspired to utilize the rate of change of the
 trace distance ``$\kappa$'' to exhibit unambiguously
 this process
 \be
 \kappa[t,\rho_{1,2}(0)]=\frac{\rmd}{\rmd t}D[\rho_1(t),\rho_2(t)].
 \ee
 Apparently, for a Markovian process the monotonically reduction
 of the trace-distance implies $\kappa\leq0$.
 The existence of $\kappa>0$ during any time intervals
 identifies the non-Markovian process.

\begin{figure}
 \begin{center}
 \includegraphics*[scale=0.7]{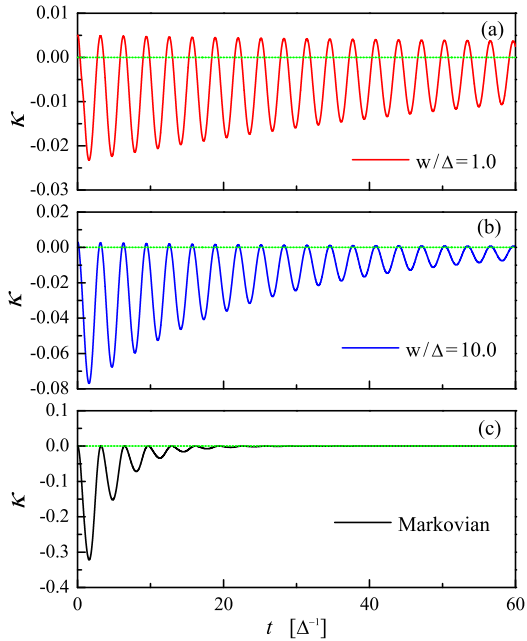}
 \caption{\label{fig5}
 The rate of change $\kappa$ of the trace distance as a
 function of time for (a) w$/\Dlt=1.0$, (b) w$/\Dlt=10.0$,
 and (c) Markovian WBL result.
 The initial pair of states used are $\rho_1(0)=|a\ra\la a|$
 and $\rho_2(0)=|b\ra\la b|$.
 The other parameters are the same as those in \Fig{fig4}.}
 \end{center}
 \end{figure}

 The numerical results are plotted in \Fig{fig5}.
 For a small  cutoff energy (w$/\Dlt=1.0$) as shown in \Fig{fig5}(a),
 there exist certain times in which $\kappa>0$.
 In those regimes, the information flows from the environment
 back to the reduced system, i.e. the non-Markovian process. It
 explains the suppression of the dephasing rate in \Fig{fig4}(d).
 An increase in cutoff energy reduces reservoir correlation time,
 and thus leads to the inhibition of the non-Markovianity
 [see \Fig{fig5}(b) for w$/\Dlt=10.0$].
 While in Markovian processes, measurements tend to wash out
 more and more characteristic features of the two states, resulting
 thus in an uncovering of these features.
 The rate of change of the trace distance ``$\kappa$'' stays
 below zero, as shown in \Fig{fig5}(c).
 The suppression of the dephasing rate due to non-Markovian
 dynamics has an important impact on noise characteristics of
 the measurement, which will be discussed
 in the next section.

 \section{\label{thsec4}Noise Characteristics}

 In this section, we first introduce the ``$N$''-resolved
 non-Markovian master equation for the calculation of the 
 output noise characteristics of the QPC detector.
 Next, the numerical results for QPC noise are presented,
 with special emphasis on the measurement SNR under 
 various conditions.

 \subsection{Particle-Number-Resolved Master Equation}

 To achieve the description of the output characteristics, the reduced
 density matrix $\rho(t)$ is unraveled into components $\rho^{(N)}(t)$, in
 which ``$N$'' is the number of electrons passing though the
 QPC during the time span $[0,t]$. The resultant time-nonlocal
 ``$N$''--resolved quantum master
 equation reads \cite{Mak01357,Fli08150601,Agu04206601,Bra06026805,Jin11053704}
 \begin{widetext}
 \begin{align}\label{CQME}
 \dot{\rho}^{(N)}(t)=-\rmi {\cal L}\rho^{(N)}(t)-\!\int_0^t\! \rmd \tau
 \bigg\{ \Pi_0(t-\tau)\rho^{(N)}(\tau)
 -\sum_{\pm}\Pi_{\pm}(t-\tau)\rho^{(N\pm1)}(\tau)\bigg\}+\varrho^{(N)}(t),
 \end{align}
 with
 \bsube
 \begin{gather}
 \Pi_0(t-\tau)(\cdots)=\sum_{ss'}\{C_{ss'}(t-\tau)Q_sG(t-\tau)Q_{s'}(\cdots)
 +[C_{ss'}(t-\tau)]^\ast G(t-\tau)(\cdots)Q_{s'}Q_{s}\},
 \\
 \Pi_\pm(t\!-\!\tau)(\cdots)
 =\sum_{ss'}\{C^{(\pm)}_{ss'}(t-\tau)G(t-\tau)Q_{s'}(\cdots)Q_s
 +[C^{(\pm)}_{ss'}(t-\tau)]^\ast Q_{s}G(t\!-\!\tau)(\cdots)Q_{s'}\}.
 \end{gather}
 \esube
 \end{widetext}
 Here, the memory kernel $\Pi_0$ corresponds to ``continuous''
 evolution of the system, and $\Pi_\pm$
 denotes forward and backward jumps of the transfer of an
 electron from the left electrode to the right one.
 By summing up \Eq{CQME} over all possible electron numbers
 ``$N$'', one straightforwardly recovers the unconditional
 master equation (\ref{QME}).
 Hereafter, we assume that the system evolves from $t_0=-\infty$, such that
 the electronic occupation probabilities at $t=0$, where electron counting
 begins, have reached the stationary state, i.e.,
 $\rho^{(N)}(t=0)=\dlt_{N,0}\rho_{\rm st}$,
 with $\rho_{\rm st}=\rho(t\rightarrow\infty)$.
 The effects of the memory of its history prior to time
 $t=0$ are incorporated in the inhomogeneity $\varrho^{(N)}$ \cite{Ema11085425}.

 The unraveling of the density matrix in \Eq{CQME} enables
 us to evaluate the probability distribution for the number
 of transferred charge $P(N,t)=\tr \{\rho^{(N)}(t)\}$,
 where the trace is over degrees of freedom of the reduced system.
 In principle, all the cumulants of the current distribution
 can be obtained, consisting thus a spectrum of full counting statistics.
 For instance, the first cumulant is directly related to the average current
 through the QPC, $I(t)=\sum_N N \dot{P}(N,t)$. By using \Eq{CQME},
 the current is given by
 \bea
 I(t)=\int_0^t \rmd \tau\, \tr\{[\Pi_-(t-\tau)-\Pi_+(t-\tau)]\rho(\tau)\}.
 \eea
 The stationary current thus reads
 \be\label{cur}
 \bar{I}\equiv
 I(t\rightarrow\infty)=\tr\{J_-(z)\rho_{\rm st}\}|_{z\rightarrow0},
 \ee
 with
 \bea
 J_\pm(z)=\tilde{\Pi}_-(z)\pm\tilde{\Pi}_+(z).
 \eea
 Here  $\tilde{\Pi}_0(z)$ and $\tilde{\Pi}_\pm(z)$, the resolvents
 of the corresponding kernels in \Eq{CQME}, are obtained by performing
 the Laplace transform
 \bsube
 \begin{align}
 \!{\ti \Pi}_0(z)(\cdots)=\sum_{ss'}\Big\{&Q_s\,\tilde{{\cal Q}}_{ss'}(z+\rmi{\cal L})Q_{s'}(\cdots)
 \nonumber \\
 &+\tilde{{\cal Q}}_{ss'}(z^\ast-\rmi{\cal L})(\cdots)\,Q_{s'}Q_s\Big\},
 \\
 \!{\ti \Pi}_\pm(z)(\cdots)=\sum_{ss'}\Big\{&\tilde{{\cal Q}}^{(\pm)}_{ss'}(z+\rmi{\cal L})Q_{s'}(\cdots)Q_s
 \nonumber \\
 &+\!Q_s\tilde{{\cal Q}}^{(\pm)}_{ss'}(z^\ast-\rmi{\cal L})(\cdots)Q_{s'}\Big\},
 \end{align}
 \esube
 where ${\cal Q}_{ss'}={\cal Q}_{ss'}^{(+)}+{\cal Q}_{ss'}^{(-)}$, with
 \begin{align}\label{Cssw}
 \tilde{\mathcal{Q}}_{ss'}^{(\pm)}(z)\equiv
 \int_{0}^{\infty}\rmd t C_{ss'}^{(\pm)}(t)e^{-z t}.
 \end{align}
 In the limit $z\rightarrow\rmi\omg$, it can be further
 simplified to
 \begin{align}
 \tilde{\mathcal{Q}}_{ss'}^{(\pm)}(z)|_{z\rightarrow\rmi\omg}
 = \tilde{C}^{(\pm)}_{ss'}(\omg)+\rmi \tilde{D}^{(\pm)}_{ss'}(\omg).
 \end{align}
 The first term denotes the coupling spectral function
 \be\label{Coupl-spec}
 \tilde{C}^{(\pm)}_{ss'}(\omg)\equiv \int_{-\infty}^{\infty}
 \rmd t C_{ss'}^{(\pm)}(t)e^{-\rmi\omg t},
 \ee
 which is associated with particle transfer processes, with
 interactions between the qubit and QPC being properly
 accounted for.
 For a Lorentzian band structure [see \Eq{Jw}], it can be evaluated
 explicitly as
  \begin{align}\label{Cpm}
 \tilde{C}_{ss'}^{(\pm)}(\omg)=&\frac{\eta\chi_s\chi_{s'}}{e^{\beta (\omg\pm V)}-1}
 \frac{4\rmw^2}{(\omg\pm V)^2+4\rmw^2}\bigg\{\frac{\rmw}{2}\varphi(\omg\pm V)
 \nonumber \\
 &+\frac{\rmw^2}{\omg\pm V}[\phi(\omg\pm V)-\phi(0)] \bigg\},
 \end{align}
 where $\eta=2\pi\Gam_{\rm L}^0\Gam_{\rm R}^0$,
 $\phi(x)$ and $\varphi(x)$ denote the real and imaginary
 parts of the digamma function
 $\Psi(\frac{1}{2}+\beta\frac{\rmw+\rmi x}{2\pi})$, respectively.
 Note here due to finite cutoff energy of the QPC detector and quasistep
 feature in the Fermi functions in \Eq{CCF}, $\tilde{\mathcal{Q}}_{ss'}^{(\pm)}(\omg)$
 decays exponentially when $\omg$ goes beyond the cutoff energy. As a result, the
 resolvents of the kernels $\tilde{\Pi}_0(z)$ and $\tilde{\Pi}_\pm(z)$ vanish in the
 limit $\omg\rightarrow\infty$.
 It should also be stressed that the present spectrum functions satisfy
 the detailed--balance relation, i.e.
 $\tilde{C}_{ss'}^{(+)}(\omg)=e^{-\beta(\omg+V)}\tilde{C}_{ss'}^{(-)}(-\omg)$,
 which means that our approach properly accounts for the energy
 exchange between the qubit and the detector during
 measurement.
 This is the reason we get nonzero stationary value for the
 real part of the off-diagonal matrix element $\rho_{ab}$,
 in contrast to that obtained in Ref. \onlinecite{Gur9715215}.

 With the Knowledge of the spectral function, the dispersion function
 $\tilde{D}_{ss'}^{(\pm)}(\omg)$
 can be obtained via the Kramers-Kronig relation
 \be \label{dispersion}
 \tilde{D}_{ss'}^{(\pm)}(\omg)=\frac{1}{\pi}\,{\cal P}
 \!\!\int_{-\infty}^{\infty} d\omg'
 \frac{\tilde{C}_{ss'}^{(\pm)}(\omg')}{\omg-\omg'},
 \ee
 where ${\cal P}$ stands for Cauchy's principal value.
 Physically, the dispersion accounts for the coupling-induced energy
 renormalization of the internal
 energies \cite{Xu029196,Yan05187,Cal83587,Wei08}.

 The second cumulant of the current distribution corresponds to
 the shot noise. To study the finite-frequency spectrum, we employ
 the MacDonald's formula \cite{Mac62}
 \be \label{MacD}
 S(\omg)=2\omg\int_0^\infty \rmd t\sin(\omg t)
 \frac{\rmd}{\rmd t}[\la N^2(t)\ra-(\bar{I}t)^2],
 \ee
 with $\la N^2(t)\ra\equiv\sum_NN^2P(N,t)$. By utilizing \Eq{CQME},
 it is simplified to
 \begin{align}\label{Sw}
 S(\omg)=S_0+4\omg {\rm Im}[\tr\{J_-(z)\tilde{N}(z)\}]|_{z\rightarrow\rmi\omg},
 \end{align}
 where the noise pedestal $S_0=2\tr\{J_+(0){\rho}_{\rm st}\}$ is
 the shot noise of the QPC detector alone.
 In the WBL and large voltage, it reproduces
 the well-known result $S_0=2e\bar{I}$ \cite{Kor995737}.
 Rich information about qubit measurement dynamics is contained
 in the excess noise (second term) in \Eq{Sw}.
 Here, $\tilde{N}(z)$ is the Laplace space counterpart of
 $N(t)\equiv\sum_N N\rho^{(N)}(t)$. By employing the ``$N$''-resolved
 quantum master equation (\ref{CQME}), it can be solved from the
 following algebraic equation
 \bea\label{Nz}
 z\tilde{N}(z)=-\rmi{\cal L}\tilde{N}(z)-\tilde{\Pi}(z)\tilde{N}(z)
 +\frac{J_-(0)\tilde{\rho}_{\rm st}}{z},
 \eea
 with $\tilde{\Pi}(z)\equiv\tilde{\Pi}_0(z)-\tilde{\Pi}_+(z)-\tilde{\Pi}_-(z)$.

 \subsection{Noise characteristics}

 The basic physics of the measurement process is the trade--off
 between acquisition of information about the state of the qubit
 and backaction dephasing of this system.
 For a quantum-limited detector, the rates of the two processes
 coincide, while for a less efficient detector, the qubit dephasing
 is more rapid than information acquisition.
 It imposes a fundamental limit on the SNR for
 a weakly measured qubit, known as the K-A bound \cite{Kor01165310}.
 An interesting feature is that the K-A bound is closely related to
 oscillation peak (around the hybridization energy $\Dlt=\sqrt{\epl^2+(2\Omg)^2}$)
 in the noise spectrum of the QPC detector, i.e.
 the maximum peak height can reach 4 times larger
 than the noise pedestal for a quantum-limited detector.

 To see how this bound emerges, let us first briefly
 derive this inequality for the Markovian case.
 We start with the current-correlation function
 $K(t)=\la \hat{I}(t+\tau)\hat{I}(t)\ra_{t\rightarrow\infty}$,
 where the average is taken over the whole system.
 The evolution of the QPC current operator $\hat{I}(t)$ is determined
 by the entire system Hamiltonian, which yields \cite{Kor01165310}
 \be
 K(\tau)=e\bar{I}\delta(\tau)
 +\frac{(\delta I)^2}{4}\tr[\sgm_z\sgm_z(\tau)\rho_{\rm st}],
 \ee
 with $\bar{I}$ the stationary current and  $\rho_{\rm st}$
 is the steady state of the qubit.
 The tr[$\cdots$] denotes the trace over the state of the
 reduced system.
 The current change $\delta I=I_a-I_b$ reflects the current
 response to electron oscillations between the dots, where
 $I_{a} (I_b)$ corresponds to the QPC current
 when the electron occupies the state $|a\ra(|b\ra)$.
 Apparently, $K(\tau)$ reflects the correlation function of
 the electron position in the dots given by $\sgm_z$.
 The evolution of the $\sgm_z(\tau)$ can be found by
 expanding the evolution operator of the entire system to
 the second order in the coupling constant, and then averaging
 over the reservoir states to obtain the equations of motion,
 with dephasing rate given by $\Gam_\rmd$.
 In the case of $\epl=0$, the noise spectrum is given
 by \cite{Kor01165310}
 \be
 S(\omg)=S_0+\frac{(\delta I)^2\Gam_\rmd\Dlt^2}{(\omg^2-\Dlt^2)^2
 +\Gam_\rmd^2\omg^2},
 \ee
 where $S_0=2e\bar{I}$ is the output noise of the QPC detector
 alone, i.e. the noise pedestal.
 At the qubit oscillation frequency $\omg=\Dlt$, the noise
 spectrum has a maximum ``signal'' of $(\delta I)^2/\Gam_\rmd$.
 The SNR thus is limited:
 \be
 {\rm SNR}\equiv\frac{S(\Dlt)-S_0}{S_0}\leq4.
 \ee
 This is the K-A bound. It has been confirmed in
 Refs. \onlinecite{Goa01235307,Rus03075303,Shn04840},
 generalized in Refs. \onlinecite{Mao04056803,Mao05085320},
 and measured in Ref. \onlinecite{Ili03097906}.
 However, several schemes have been proposed recently to overcome
 the K-A bound, which can be divided into two categories.
 The first one concerns with increasing the signal, such
 as quantum nondemolition measurements \cite{Ave02207901,Jor05125333}
 and quantum feedback control \cite{Wan07155304,Vij1277}.
 The second type is to reduce the pedestal noise by employing a strongly
 responding SET \cite{Jia09075320} or twin detectors \cite{Jor05220401}.
 In this work, we find the non-Markovian processes allow
 a violation of the K-A bound on the SNR.
 The details together with an interpretation will be provided
 later.

 \begin{figure}
 \begin{center}
 \includegraphics*[scale=0.78]{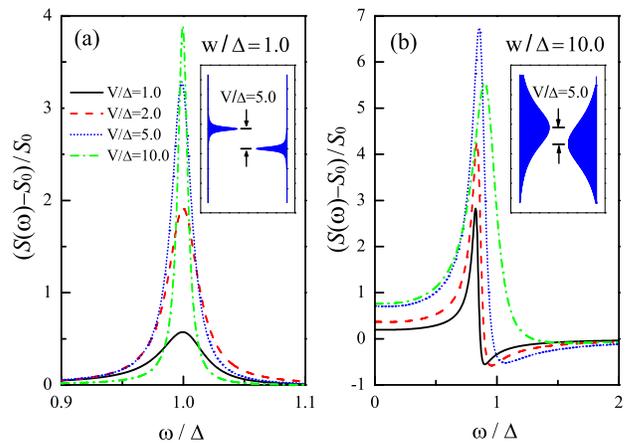}
 \caption{\label{fig6}
 Noise feature for a symmetric qubit ($\epl=0$) under different measurement
 voltages for (a) w$/\Dlt=1.0$ and (b) w$/\Dlt=10.0$.
 The inset in (a) and (b) shows the reservoir's
 spectral density for corresponding cutoff energies and fixed measurement voltage
 $V/\Dlt=5.0$.
 The temperature and other parameters are the same as those in \Fig{fig4}.}
 \end{center}
 \end{figure}

 The computed noise is shown in \Fig{fig6} for different
 cutoff energies and voltages.
 For a small cutoff energy (w$/\Dlt=1.0$), prominent non-Markovian
 processes take place, which leads to a strongly suppressed
 dephasing rate [cf. \Fig{fig4}(d)].
 It is reflected in the noise spectrum as the
 narrow width of the oscillation peak.
 As the measurement voltage increases, the qubit will be excited,
 which leads to the rising peak height (``signal'') and SNR.
 However, the SNR cannot exceed the limit of ``4'', even in the
 limit of $V/\Dlt\rightarrow\infty$, as we have verified.

 In the case of large cutoff energy (w$/\Dlt=10.0$), however,
 violation the K-A bound is observed unambiguously [see,
 for instance, the dotted curve in \Fig{fig6}(b)].
 The violation is due to the presence of non-Markovian processes,
 in which reversed flow of information from the environment
 back to the reduced system takes place.
 This mechanism is analogous to the quantum feedback
 scheme \cite{Wan07155304,Vij1277}. Yet, there, one has to
 implement an extra procedure in which
 the measurement information
 in the detector is converted into the evolution of a qubit state.
 Our analysis thus serves a direct and transparent way to improve
 the efficiency in quantum measurement.

 One may ask why the K-A bound is not violated in the case of
 a small cutoff (w$/\Dlt=1.0$), where prominent non-Markovian
 processes are present.
 This is actually associated with the energy that needed to
 excite the qubit.
 Let us consider the situation of a voltage $V/\Dlt=5.0$.
 For a small cutoff (w$/\Dlt=1.0$), the number of channels for electrons
 to tunnel though the detector is remarkably suppressed, see the
 schematic density spectral in the inset of \Fig{fig6}(a).
 It restricts the number of electrons that can provide
 energy to excite the qubit, and eventually results in
 the SNR below the limit of ``4''.
 Unlike the cases of w$/\Dlt=1.0$, the number of channels
 are considerably increased for a large cutoff energy (w$/\Dlt=10.0$),
 as shown in the inset of \Fig{fig6}(b).
 Therefore, sufficient energy is provided to excite the
 qubit, which leads eventually to the violation of the K-A bound.
 However, even in the case of large cutoff energy, the
 number of channels does not necessarily increase with
 rising voltage.
 For instance, there are less effective channels in the case
 of $V/\Dlt=10.0$ than that of $V/\Dlt=5.0$.
 One thus observes a suppressed SNR for $V/\Dlt=10.0$
 in comparison with that of $V/\Dlt=5.0$, as shown by the
 dot-dashed curve in \Fig{fig6}(b).

 Furthermore, the qubit-QPC coupling gives rise to a
 dynamical renormalization of the qubit energies
 [see \Eq{dispersion}], which vanishes at ``w$\rightarrow0$ and
 increases with the cutoff \cite{Luo09385801}.
 On one hand, it leads to the shift of the oscillation peaks
 towards low frequencies, as shown in \Fig{fig6}(b).
 One the other hand, the energy renormalization gives rise to
 incoherent jumps between the two states.
 The detector attempts to localize the electron in one of the
 states for a longer time, leading thus to the quantum Zeno effect,
 which is manifested as the non-zero noise at zero frequency,
 as displayed in \Fig{fig6}(b).

 \begin{figure}
 \begin{center}
 \includegraphics*[scale=0.78]{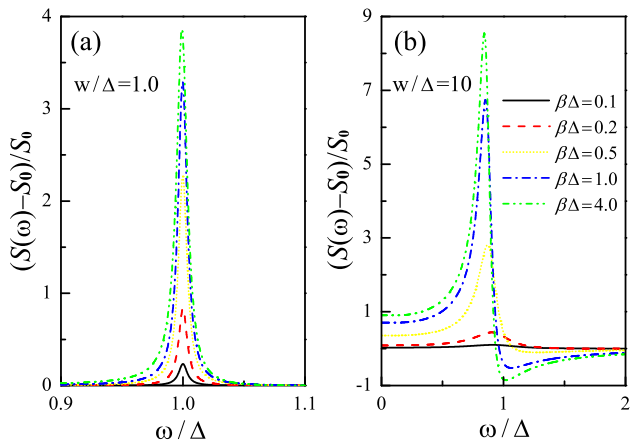}
 \caption{\label{fig7}
 Noise spectrum of a symmetric qubit at various temperatures
 and cutoff energies (a) w$/\Dlt=1.0$ and (b) w$/\Dlt=10.0$.
 The measurement voltage is $V/\Dlt=5.0$.
 The other parameters are the same as those in \Fig{fig4}.}
 \end{center}
 \end{figure}

 We are now in a position to discuss the influence of the temperature
 on the noise spectrum. The numerical results are plotted in
 \Fig{fig7}.
 For a small cutoff energy (w$/\Dlt=1.0$), the
 presence of a prominent non-Markovian effect inhibits
 the dephasing rate, which is insensitive to the temperature, as
 shown in \Fig{fig3}(b).
 The width of the oscillation peak thus is strongly suppressed
 for all the temperatures.
 Furthermore, it is found that the height of the oscillation peak
 decreases rapidly with rising temperature.
 The reason is attributed to the enhanced qubit relaxation rate
 as the temperature grows, analogous to the finding in the
 the Markovian limit \cite{Li05066803}.
 In this regime of small cutoff energy, the SNR cannot exceed
 the K-A bound as we have checked.
 It is again owing to the limited number of channels that electrons
 can transfer through the QPC.
 However, in the case of large bandwidth w$/\Dlt=10.0$ strong
 violation of the K-A bond is observed at a low temperature
 $\beta\Dlt=4.0$ [see \Fig{fig7}(b)].
 As the temperature grows, the oscillation peak is again
 reduced due to qubit relaxation, similar to the situation
 of w$/\Dlt=1.0$.

 To complete this section, we discuss the situation under which
 the violation of the K-A bound may take place.
 In the case of small cutoff energy (w$<\Dlt$), the SNR cannot
 exceed the limit of ``4'' under arbitrary voltage and temperature,
 even though there is a strong non-Markovian effect.
 It is ascribed to the limited number of channels that electrons
 can transfer though the detector.
 It cannot provide enough energy to fully excite the qubit,
 thus restricts the ``efficiency'' of the measurement.
 In this sense, a small cutoff energy works as a suppression mechanism
 to the ``effectiveness'' of the voltage and temperature.
 In the opposite WBL (w$\rightarrow\infty$),
 one expects very short reservoir correlation time, approaching
 thus to the Markovian case.
 Our result reproduces to the previous Markovian ones.
 In this case, the information flows purely from the reduced
 system to the reservoir and the SNR is limited to 4.
 Therefore, the violation of the K-A bound only occurs for
 moderate cutoff energies, together with an appropriately
 large measurement voltage and a low temperature.
 In this regime, enough energy will be provided to excite
 the qubit, while relaxation to the ground state takes places
 slowly.
 Moreover, the presence of finite non-Markovian dynamics
 results in the opportunity for the information to flow
 from the reservoir back to the system, which eventually
 lead to an SNR exceeding the K-A bound.
 In comparison with the quantum feedback scheme, the present
 work serves as a straightforward and transparent way to
 improve the ``efficiency'' in quantum measurement.

 \section{\label{thsec5}Conclusions}

 In summary, we have investigated the dynamics of a charge qubit
 under continuous measurement by a quantum point contact, with
 special attention paid to the non-Markovian measurement characteristics.
 We identified the regimes where prominent non-Markovian memory effects
 are present by analyzing how the reservoir's correlation time is
 varied as function of different parameters of the QPC detector.
 In comparison with the Markovian case, considerable differences in
 qubit relaxation and dephasing behaviors were observed in the
 non-Markovian domain.
 Furthermore, the non-Markovian dynamics was found to have a vital
 role to play in the output noise features of the detector.
 In particular, we observed unambiguously that the signal-to-noise
 ratio can exceed the limit of ``4'', leading thus to the violation
 of the Korotkov-Averin bound.
 In comparison with other approaches, such as quantum feedback
 scheme, our results might open new possibilities to enhance
 measurement efficiency in a straightforward and transparent
 way.

\begin{acknowledgments}
 Support from the National Natural Science
 Foundation of China (11204272, 11147114, and 11004124),
 the Zhejiang Provincial Natural Science Foundation
 (Y6110467 and LY12A04008) is gratefully acknowledged.
\end{acknowledgments}


\end{document}